# Groundwater pumping to increase food production causes persistent groundwater drought in India

Akarsh Asoka[1] and Vimal Mishra[1]
1. Civil Engineering and Earth Sciences, Indian Institute of Technology (IIT) Gandhinagar, Gujarat, India

Corresponding author: vmishra@iitgn.ac.in

**Abstract**

Rapid groundwater depletion in India is a sustainability challenge. However, the crucial role of climate and groundwater pumping on persisting groundwater drought remains unrecognized. Using the data from Gravity recovery climate experiment (GRACE) satellites and more than 5000 observational wells, here we show that the increase in precipitation in northwest India (NWI) no longer helps to recover from groundwater drought that started after 2012. Groundwater storage anomaly (GWSA) from the GRACE well observations is strongly linked with accumulated precipitation for 153, 105, and 18 months for NWI, northcentral (NCI), and south India (SI). Precipitation and GWSA have decoupled in NWI after 2012 indicating the higher influence of groundwater pumping for crop production than climate. The relative contribution of vegetation growth ($R^2$=0.26) on GWSA is higher than precipitation ($R^2$=0.02) for 2002-2016 in NWI than in NCI and SI. Our findings highlight the urgent need of reducing groundwater pumping in India.

**Introduction**

Groundwater remains the lifeline of human survival in India, and unsustainable groundwater pumping is the primary cause of groundwater depletion (Döll et al., 2012; Matthew Rodell et al., 2009; Taylor et al., 2013). Groundwater is crucial during droughts to meet drinking and irrigation requirements (Wada et al., 2010). India is the largest user of groundwater where more than 60% of the agricultural area is irrigated from groundwater. Groundwater pumping in India has increased exponentially after the introduction of mechanical pumps and subsidized power supply (Mishra et al., 2018b; T. Shah, 2009). Excessive abstraction of groundwater for irrigation in India resulted in groundwater depletion in many regions with the most prominent decline in northwestern India (Asoka et al., 2017, 2018; Matthew Rodell et al., 2009; Tiwari et al., 2009). Additionally, groundwater storage in India has been affected by the declining monsoon precipitation (Asoka et al., 2017) and the changing character of precipitation (Asoka et al., 2018). Low-intensity precipitation that is favourable to groundwater recharge in north India has significantly declined during 1951-2016 (Asoka et al., 2018).

Unsustainable groundwater pumping for irrigation combined with declining precipitation can cause a rapid groundwater depletion and groundwater drought. Nonrenewable groundwater abstraction in India is the highest in the world, which account for 19 % (68 km$^3$/yr) of the gross irrigation demand (Wada et al., 2012). Groundwater drought, with prolonged below normal groundwater level, adversely affects the ecosystem's resilience (Bloomfield & Marchant, 2013; Lanen & Peters, 2000). Groundwater depletion in India has been widely reported (Asoka et al., 2017; Matthew Rodell et al., 2009; Tiwari et al., 2009). However, the role of climate and vegetation on groundwater drought in north India remains unrecognized. Here using the groundwater data from GRACE satellites and more than 5000 observation wells, we report the role of vegetation and climate (especially precipitation variability) on persisting groundwater drought in India.

**Data and Methods:**

**Datasets**

We obtained the monthly Terrestrial Water Storage Anomaly (TWSA) at 1° spatial resolution from Gravity Recovery and Climate Experiment (GRACE), which is available from Centre for Space Research (CSR Version 5) for 2002 to 2016. We derived the monthly groundwater storage anomaly (GWSA) by subtracting surface water storage anomaly from GRACE TWSA. We used the total of soil moisture, canopy storage, and snow water equivalent to representing surface water storage from the four land surface models (LSMs: Noah, CLM, VIC, and MOSAIC) that are available from the Global Land Data Assimilation System (GLDAS) (M. Rodell et al., 2004). For a detailed methodology to estimate GRACE GWSA, please refer to Asoka et al.,(2017) . Additionally, we used below ground level (m bgl) water table observations from the Central Groundwater Board (CGWB) for the period 1996-2016. GRACE GWSA represents both shallow and deep groundwater while the majority of well observations from CGWB are from shallow wells (Mishra et al., 2018a). We obtained water table observations from more than 20000 wells that have data for January, May, August, and November. Since the well observations have quality issues related to data gaps and other inconsistencies, after scrutiny, we selected 5874 wells with consistent long-term records for further analysis as in (Asoka et al., 2018). We estimated groundwater storage anomaly by multiplying monthly anomalies of well level observations with the specific yield as in (Asoka et al., 2017 and Asoka et al. (2018). We used the specific yield values of 14 major aquifers recommended by CGWB based on long duration pumping tests (Asoka et al., 2018).

Apart from groundwater well observations, we used gridded daily precipitation from India Meteorological Department (IMD), which has been developed using observations from more than 6000 gage stations and inverse distance weighting method (Pai et al., 2014). Gridded precipitation from IMD is widely used for hydroclimatic studies in India (Mishra et al., 2014; H. L. Shah & Mishra, 2016). In our analysis, we first identified an optimal period for which groundwater storage variability is strongly coupled with the historically accumulated precipitation. We estimated the optimal period for which correlation between GWSA from GRACE or well data and accumulated precipitation anomaly was the highest. The optimal period provides us an idea about how the past climate is linked with groundwater storage. For instance, a high optimal period means that long term precipitation is linked with the groundwater storage variability. On the other hand, a low optimal period suggests that the groundwater in a given region is not strongly coupled with the long-term precipitation; rather it has a shorter persistence.

**Estimation of Optimal Period**

We used accumulated precipitation for 1-180 months (about the past 15 years) to estimate the optimal period for GWSA. We estimated the correlation between accumulated precipitation and GRACE GWSA for each grid for the accumulated time-period that varies from 1 to 180 months.

To estimate the optimal period, we used a median correlation approach in our analysis, which is robust against the extremes. To do so, we started the correlation analysis with the initial 5-year time series (2002-2006: 60 data points) of GWSA and accumulated precipitation. We then estimate the correlation between GWSA and accumulated precipitation by adding monthly data (GWSA and precipitation) from 2006 to 2016. Therefore, instead of estimating the correlation coefficient for the entire time (2002-2016), we estimated correlation for various periods and then took the median correlation coefficient. This way the correlation is not affected by extreme values of precipitation and GWSA. We performed the analysis for each accumulation period (1-180 month) of precipitation. We obtained the median correlation (r-value) and significance level (p-value) for each accumulation period (from121 correlation values for each accumulation period of precipitation and the median values were chosen). The time-period with maximum (+ve) correlation (at 95 % confidence level) is assigned as the optimal period. For the regional analysis, we used the area-averaged time series of GWSA and precipitation for northwest India (NWI), north-central India (NCI) and south India (SI). We performed a similar analysis using GWSA derived from well level observations, which mostly represents shallow (depth 20m or less) water table. We started with 1996-2005 time series ( 40 data points: 4 observations per year) and serially added the data up to 2016. The median correlation of each accumulation period was used to estimate the optimal period. We assigned precipitation for each well from its nearest neighbouring grid cell.

**Estimation of groundwater drought**

We identified the groundwater drought based on GWSA using the methodology described in Thomas et al., (2014). Groundwater drought is identified if the departure of GWSA is below normal for at least three consecutive months (Thomas et al., 2014).  For the continuity of data for drought analysis, we filled the missing data using linear interpolation. Here, we should note that groundwater drought can be a manifestation of both climatic and anthropogenic factors as reported in previous studies (Asoka et al., 2017; Matthew Rodell et al., 2009; Tiwari et al., 2009). We used GWSA after removing monthly climatology (2002-2016)) to identify the groundwater drought period.

**The linkage between irrigation and groundwater**

We used Advance Very High-Resolution Radiometer (AVHRR) no noise (smoothed) Normalized Difference Vegetation Index (NDVI) from 1982-2016 to evaluate the role of irrigation on groundwater storage variability. We obtained a weekly composite of NDVI at 4 km spatial resolution from National Ocean Atmospheric Administration (NOAA), which is calibrated and validated to estimate phenological phase and start and end of the season (https://www.star.nesdis.noaa.gov/smcd/emb/vci/VH/vh_browse.php). We aggregated weekly NDVI to monthly means and then gap filled using linear interpolation (Knauer et al., 2016). We used land use land cover (LULC) data obtained from the National Remote Sensing Centre (NRSC) at a 56m spatial resolution to mask out nonagricultural areas. The LULC data is derived using Advanced Wide Field Sensor (AWiFs) for 2007–2008. We spatially aggregated the 56 m LULC data to 4km resolution (same as NDVI resolution) based on majority resampling method.

We used Food and Agriculture Organization (FAO) Global Map of Irrigated Area (GMIS) version 5 available at 0.083° resolution (Siebert et al., 2013) to estimate the role of irrigation on NDVI and groundwater storage variability in India. We identified non-irrigated and irrigated with groundwater pixels using GMIS data. We assigned fractional area (%) irrigated with groundwater for each 4km grid-cell using nearest neighbourhood (more than 60 % area irrigated with groundwater). We finally spatially aggregated 4km NDVI irrigated with groundwater to 1° to make it consistent with the GRACE dataset. We aggregated GWSA and NDVI for the three regions (NWI, NCI, and SI) for irrigated with groundwater. Finally, We estimate the correlation between GWSA and 4, 12, and 24 months accumulated NDVI anomalies to evaluate the role of irrigated agriculture on groundwater storage variability in India. We performed the correlation analysis between NDVI for Kharif (JJAS) and Rabi (ONDJFM) seasons and GWSA for irrigated areas to estimate seasonal variability in irrigation impacts on groundwater storage variability in the three regions. We also compared the magnitude of NDVI (spatially averaged for NWI, NCI, and SI) in the groundwater irrigated area and non-irrigated (less than 20% of the total area equipped for irrigation) areas for the Kharif and Rabi seasons.

To assess the importance of precipitation and irrigated agriculture on groundwater storage variability, we estimated relative importance (Asoka et al., 2018; Silber et al., 1995) of precipitation and NDVI on GWSA. We estimated the relative contribution (as the coefficient of determination ($R^2$)) of precipitation for an optimal period and 4, 12 and 24 months averaged NDVI (groundwater irrigated) on groundwater storage anomaly of NWI, NCI, and SI. The analysis is performed at 95 % confidence level for 1000 bootstrap runs. More details on relative contribution can be obtained from (Asoka et al., 2018).

Results

**Optimal Period of Groundwater Storage Variability**

We start our analysis by estimating the optimal period (Bloomfield & Marchant, 2013; Kumar et al., 2016), which is an accumulation period of precipitation at which groundwater storage anomaly shows maximum correlation (see methods for more details, Figure. 1). Groundwater recharge is a slow process and can take many years to show the influence of precipitation. Therefore, we considered 1-180 months accumulated precipitation to estimate the optimal period. Since the GRACE groundwater storage anomalies reflect the influence of climate on both deep and shallow aquifers, we first estimate the optimal period based on the GRACE data (2002-2016). Then, we use groundwater well data from more than 5000 wells from the Central Groundwater Board (CGWB), which are available for a relatively longer period (1996-2016) than the satellite-based estimates.

We estimated correlation coefficient (r) between monthly GRACE groundwater storage anomaly and accumulated precipitation for different (1-180 months) accumulation periods (Figure. 1). We find that a majority of north India shows a significant (p-value < 0.05) positive correlation between GRACE groundwater storage anomaly and accumulated precipitation at relatively higher (more than eight years) accumulation period (Fig 1. A&B). There is a large spatial variability in accumulation period in the northwestern India (NWI), which is attributable to higher groundwater abstraction for pumping (Matthew Rodell et al., 2009; Tiwari et al., 2009).

The localized and excessive groundwater pumping (Asoka et al., 2017; Mishra et al., 2018a; Matthew Rodell et al., 2009; Tiwari et al., 2009), aquifer characteristics (specific yield, size, depth of groundwater table, and persistence) and precipitation variability may contribute to higher optimal period in NWI and in north-central India (NCI).

Area averaged groundwater storage anomalies from GRACE satellites reveal the highest optimal period (153 months; around 13 years) for NWI, followed by NCI (105 months; around nine years) and SI (18 months; around 1.5 years). We find a strong and significant (p-value <0.05) coupling between GRACE groundwater storage anomalies and accumulated precipitation for the optimal period for NWI (correlation (r)=0.66±0.26), NCI (r=0.85±0.18), and SI (r=0.78±0.04) [Figure. 1, Table 1]. Relatively lower correlation between groundwater storage anomaly and accumulated precipitation in NWI may be due to higher anthropogenic influence (i.e., groundwater pumping) on groundwater storage and due to the decoupling of accumulated precipitation and groundwater storage anomaly after 2012 (Figure. 1C). This decoupling of precipitation and groundwater storage anomaly shows the influence of excessive groundwater pumping (Mishra et al., 2018) in NWI (Figure. 1C). In contrast, we note a strong linkage between accumulated precipitation and groundwater storage anomaly for NCI and SI (Figure. 1D, E), where the influence of groundwater pumping is lesser than NWI. The linkage between accumulated precipitation and groundwater storage for NCI and SI is positive for all the accumulation periods (Figure. S1) while NWI shows a negative correlation for shorter (less than eight years) accumulation periods. The negative correlation between groundwater storage and precipitation for a short accumulation period indicates that depleted groundwater storage in NWI needs a longer time to get replenished.

Next, we estimate the optimal period for groundwater storage obtained from more than 5000 observational wells in India (Figure. 2). One of the advantages of the well data is that it covers a longer period (1996-2016) than the GRACE estimates (2002-2016). However, as the majority of central groundwater board (CGWB) monitoring wells are shallow (less than 20m), these may underestimate the groundwater depletion due to pumping for irrigation, which happens primarily from the deep aquifers (Mishra et al., 2018a). Notwithstanding, these differences between the GRACE and well data, we find that our estimation of the optimal period is robust as both (GRACE and well) the datasets show a higher optimal period for the groundwater storage anomalies in NWI (Figure. 2). The optimal period estimated from well data for NWI, NCI, and SI is 136, 63, and 13 months, respectively (Figure. 2). Similar to the GRACE data, groundwater storage anomalies from wells show a strong and significant (p-value < 0.05) linkage between accumulated precipitation for all the three (NWI, NCI, and SI) regions, with a correlation coefficient of 0.89±0.04, 0.85±0.05, and 0.86±0.01 for NWI, NCI, and SI, respectively. The difference in the optimal period between the GRACE and well estimates is primarily due to the absence of deep wells in the CGWB record. Overall, we find that both GRACE and well data show a higher optimal period for NWI followed by NCI and SI.

A majority of NWI and NCI consists of the large capacity alluvial aquifer with higher persistence (Figure. S2), the longer optimal period in the NWI is potentially linked to excessive abstraction for irrigation (Matthew Rodell et al., 2009) and relatively lower annual precipitation as the region is in the semi-arid zone. Groundwater storage in SI shows a shorter optimal period due to smaller aquifer size and their hard rock and basaltic nature (Asoka et al., 2017). Moreover, the optimal period also depends on the persistence of groundwater storage anomaly (Figure. S2), which is linked with site and aquifer characteristics (Bloomfield & Marchant, 2013) and can be influenced by groundwater pumping. Other factors that govern the optimal period include bedrock depth, depth to water table, and average annual precipitation (Li & Rodell, 2015). For instance, a relatively longer optimal period is observed in the areas with deep groundwater level (NWI) compared to shallow (SI) bedrock regions (Li & Rodell, 2015). We show that groundwater storage variability in NWI and NCI is affected by longer-term (8-15 years) variability in precipitation along with the groundwater pumping for irrigation. In contrast, groundwater storage in SI is more strongly linked with relatively short-term (1-2 years) precipitation variability. Our results have implications for groundwater sustainability in India as in NWI and NCI it takes longer to replenish groundwater. Moreover, a significant depletion of groundwater in north India that caused persistent groundwater drought and makes the recovery of groundwater challenging due to high optimal period and groundwater pumping.

**Persistent groundwater drought in North India**

Long-term meteorological droughts reduce groundwater recharge and enhance groundwater abstraction for irrigation and the combined effect often results in groundwater drought (Kløve et al., 2014; Taylor et al., 2013). We estimated groundwater drought in India using groundwater storage anomaly from GRACE during 2002-2016 (Figure. 3). We find that the longest groundwater drought had a period of 56 (04/2012 to 12/2016), 32 (04/2014 to 12/2016), and 21 (04/2002 to 12/2003 & 02/2004 to 10/2005) months in NWI, NCI and SI (Figure. 3 A-C and Table S2). Here, we note that the groundwater drought that started in 2012 and 2014 in NWI and NCI is not demised during the observational record of 2002-2016 and may be still ongoing. Therefore, the recent groundwater drought in NWI and NCI is persistent and yet to be recovered.

Large areal coverage of groundwater drought in NWI and NCI is caused by a massive groundwater depletion (Matthew Rodell et al., 2009; Tiwari et al., 2009), which is due to combined effect of climate and groundwater pumping in north India (Figure.3 A-C). NWI and NCI experienced a prolonged and widespread groundwater drought during recent years (Figure. 3 A&B). Moreover, monthly groundwater storage anomaly for NWI has decoupled from the long-term precipitation anomaly during the recent year (Figure. 1C, after 2012 in NWI). The decoupling and a weaker relationship between groundwater storage (from GRACE) and precipitation anomalies reflect the prominence of groundwater pumping in NWI. In NCI and SI, GRACE groundwater storage and precipitation anomaly for the optimal period are strongly associated indicating a relatively lower influence of groundwater pumping (Figure. 3B, C). The recent (2015-2016) meteorological droughts caused an enormous groundwater depletion (Mishra et al., 2016) in north India as both NWI and NCI experienced the most widespread groundwater drought in December 2016 (Figure. 3D). Similarly, the most widespread groundwater drought in SI occurred in January 2003. (Figure. 3E). We find that north India has experienced the most widespread and persistent groundwater droughts during the recent years while south India did not face a major groundwater drought in the recent years (Figure. S3, Table S2).

**Role of vegetation and climate on groundwater drought**

Finally, we identify the role of vegetation and climate on groundwater drought in India. We consider vegetation growth in the groundwater based irrigated regions as a surrogate of groundwater pumping while cumulative precipitation anomaly for the optimal period to represent the role of climate (Figure. 4). We evaluate the role of vegetation on groundwater storage variability using normalized difference vegetation index (NDVI) (Figure. 4, please see methods for details). Our results show a negative correlation between 12-month NDVI and groundwater storage anomaly in the majority of NWI and part of NCI (Figure. 4A). However, the majority of SI shows a strong positive correlation between 12-month NDVI and groundwater storage anomaly (Figure. 4 D-F). Based on aggregated 12-month NDVI for NWI, NCI, and SI, we note a negative correlation for NWI (r=-0.52±0.07) and NCI (r=-0.14±0.11) while a strong positive correlation (r=0.67±0.03) in SI (Figure. 4 D-F & Table S3). This negative relationship (between 12-month NDVI and groundwater storage anomaly) in NWI and NCI may be linked to greening trends in vegetation (Akarsh & Mishra, 2015) and groundwater depletion. Both greening trends of vegetation and groundwater depletion are potentially linked with groundwater pumping for irrigation.

We separately analyzed NDVI for the two major crop growing seasons (Kharif and Rabi). We find a negative correlation (r= -0.71, p=0.003) between NDVI and groundwater storage anomaly in the majority of NWI and in parts of NCI (r=-0.34, p=0.218) and a positive correlation (r=0.27, p=0.334) for SI for the Kharif (JJAS) season (Figure. 4B, Figure. S4). Moreover, NWI shows a negative relationship (r= -0.49, p=0.074) between NDVI and groundwater storage anomaly for Rabi (ONDJFM) season indicating the role of groundwater pumping on the greening of vegetation (Figure. 4C, Figure. S4 A, D). Our results show that NDVI is considerably higher in irrigated (with groundwater) than the non-irrigated area in NWI in both Kharif and Rabi seasons for 1982-2016 (Figure. S5). Negative relationship (between groundwater and NDVI) and high NDVI in groundwater based irrigated regions of NWI show the indirect influence of groundwater pumping for irrigation on groundwater depletion. The groundwater depletion for irrigation caused persistent drought in NWI despite the positive trends in precipitation for the optimal period.

To separate the role of climate and vegetation on groundwater storage variability, we estimated the relative contribution of precipitation for optimal period (153 months: NWI, 105: NCI and 18: SI) and NDVI (4, 12 and 24 month) on groundwater storage anomaly in NWI, NCI and SI (Figure. S6 & Table S4). We performed the analysis for the 2002-2016 and 2002-2012 period to quantify the role of groundwater depletion after 2012. The relative contribution of 12-month NDVI ($R^2$=0.26) on groundwater storage anomaly in NWI is higher than precipitation ($R^2$=0.02) during 2002-2016 (Figure. S6). However, the contribution of precipitation ($R^2$=0.57) is higher than 12-month NDVI ($R^2$=0.13) if we consider the period before 2012 (2002-2012). The lower contribution of precipitation during 2002-2016 is primarily due to the decoupling of precipitation and groundwater storage anomaly after 2012. Our results indicate that the groundwater depletion and persistent drought after 2013 in NWI are linked with the groundwater pumping for irrigation. In NCI and SI, the relative contribution of precipitation is higher ($R^2$=0.66 for NCI; $R^2$= 0.35 for SI) than 12-month NDVI ($R^2$=0.00 for NCI; $R^2$= 0.20 for SI) during 2002-2016. We find a similar contribution of precipitation ($R^2$=0.70 for NCI; $R^2$= 0.35 for SI) and 12-month NDVI ($R^2$=0.13 for NCI; $R^2$= 0.27 for SI) for 2002-2012 as well (Figure. S6 & Table S4), which indicates a relatively lesser influence of vegetation on groundwater storage variability. Here, we note that the groundwater drought in NCI is primarily driven by precipitation variability while groundwater drought in NWI is linked with groundwater pumping. We also estimated the relative importance of 4 and 24-month NDVI and precipitation and found a similar pattern in relative contribution for both the period (Table S4).

Groundwater drought persisted after 2012 in NWI and after 2014 in NCI. Both the regions have intensive groundwater-based irrigation and are yet to recover from the groundwater drought. The recovery of groundwater drought depends on the groundwater recharge and groundwater pumping for irrigation. The monsoon season precipitation is the primary source of groundwater recharge in India other than canals and surface water storage in reservoirs and ponds. Groundwater recharge during the monsoon season depends on the total amount and nature of precipitation. For instance, low-intensity precipitation is strongly linked with the monsoon season groundwater recharge (Asoka et al., 2018). However, both low-intensity and total precipitation have significantly declined in a major part of north India (Asoka et al., 2018). We note that despite an increase of precipitation during the last few years, groundwater drought in NWI is not improved. Our results show that the rate of groundwater depletion in NWI is much faster than the recharge, which contributed to widespread and persistent groundwater drought in the region. The groundwater drought can be improved by reducing the groundwater pumping for irrigation as shown by the relative contribution of irrigated agriculture in the region. In contrast, groundwater drought in NCI is strongly coupled with the long-term decline in precipitation (Mishra et al., 2012; Roxy et al., 2015). However, the role of pumping on groundwater drought in NCI can not be neglected. The longer optimal period for groundwater storage in north India has considerable implications indicating that once groundwater is extensively depleted it would take a longer period to recover. Therefore, there is an urgent need for reducing groundwater pumping in north India to ensure drinking and irrigation water availability in the future.

**Acknowledgement:** Authors acknowledge the data availability from CGWB and GRACE for this work. All the datasets used in this study are publicly available and can be obtained from GRCAE: https://grace.jpl.nasa.gov/data/get-data/, CGWB: http://www.india-wris.nrsc.gov.in, and IMD: www.imd.gov.

**Reference**

Akarsh, A., & Mishra, V. (2015). Prediction of vegetation anomalies to improve food security and water management in India. *Geophysical Research Letters*, 2015GL063991. https://doi.org/10.1002/2015GL063991

Asoka, A., Gleeson, T., Wada, Y., & Mishra, V. (2017). Relative contribution of monsoon precipitation and pumping to changes in groundwater storage in India. *Nature Geoscience*, *10*(2), 109–117. https://doi.org/10.1038/ngeo2869

Asoka, A., Wada, Y., Fishman, R., & Mishra, V. (2018). Strong Linkage Between Precipitation Intensity and Monsoon Season Groundwater Recharge in India. *Geophysical Research Letters*, *45*(11), 5536–5544. https://doi.org/10.1029/2018GL078466

Bloomfield, J. P., & Marchant, B. P. (2013). Analysis of groundwater drought building on the standardised precipitation index approach. *Hydrology and Earth System Sciences*, *17*, 4769–4787.

Döll, P., Hoffmann-Dobrev, H., Portmann, F. T., Siebert, S., Eicker, A., Rodell, M., et al. (2012). Impact of water withdrawals from groundwater and surface water on continental water storage variations. *Journal of Geodynamics*, *59*, 143–156.

Kløve, B., Ala-Aho, P., Bertrand, G., Gurdak, J. J., Kupfersberger, H., Kværner, J., et al. (2014). Climate change impacts on groundwater and dependent ecosystems. *Journal of Hydrology*, *518*, 250–266. https://doi.org/10.1016/j.jhydrol.2013.06.037

Knauer, K., Gessner, U., Fensholt, R., & Kuenzer, C. (2016). An ESTARFM Fusion Framework for the Generation of Large-Scale Time Series in Cloud-Prone and Heterogeneous Landscapes. *Remote Sensing*, *8*(5), 425. https://doi.org/10.3390/rs8050425

Kumar, R., Musuuza, J. L., Van Loon, A. F., Teuling, A. J., Barthel, R., Ten Broek, J., et al. (2016). Multiscale evaluation of the Standardized Precipitation Index as a groundwater drought indicator. *Hydrology and Earth System Sciences*, *20*(3), 1117–1131. https://doi.org/10.5194/hess-20-1117-2016

Lanen, H. A. J. V., & Peters, E. (2000). Definition, Effects and Assessment of Groundwater Droughts. In *Drought and Drought Mitigation in Europe* (pp. 49–61). Springer, Dordrecht. https://doi.org/10.1007/978-94-015-9472-1_4

Li, B., & Rodell, M. (2015). Evaluation of a model-based groundwater drought indicator in the conterminous US. *Journal of Hydrology*, *526*, 78–88.

Mishra, V., Smoliak, B. V., Lettenmaier, D. P., & Wallace, J. M. (2012). A prominent pattern of year-to-year variability in Indian Summer Monsoon Rainfall. *Proceedings of the National Academy of Sciences*, *109*(19), 7213–7217. https://doi.org/10.1073/pnas.1119150109

Mishra, V., Shah, R., & Thrasher, B. (2014). Soil Moisture Droughts under the Retrospective and Projected Climate in India. *Journal of Hydrometeorology*, (2014), 2267–2292. https://doi.org/10.1175/JHM-D-13-0177.1

Mishra, V., Aadhar, S., Asoka, A., Pai, S., & Kumar, R. (2016). On the frequency of the 2015 monsoon season drought in the Indo-Gangetic Plain. *Geophysical Research Letters*, *43*(23). Retrieved from http://onlinelibrary.wiley.com/doi/10.1002/2016GL071407/full

Mishra, V., Asoka, A., Vatta, K., & Lall, U. (2018a). Groundwater Depletion and Associated $CO_2$ Emissions in India. *Earth's Future*. https://doi.org/10.1029/2018EF000939

Mishra, V., Asoka, A., Vatta, K., & Lall, U. (2018b). Groundwater depletion and associated $CO_2$ emissions in India. *Earth's Future*, *6*(12), 1672–1681.

Pai, D. S., Sridhar, L., Rajeevan, M., Sreejith, O. P., Satbhai, N. S., & Mukhopadhyay, B. (2014). Development of a new high spatial resolution (0.25 degrees x 0.25 degrees) long period (1901-2010) daily gridded rainfall data set over India and its comparison with existing data sets over the region. *MAUSAM*, *65*(1), 1–18.

Rodell, M., Houser, P. R., Jambor, U. e a1, Gottschalck, J., Mitchell, K., Meng, C. J., et al. (2004). The global land data assimilation system. *Bulletin of the American Meteorological Society*, *85*(3), 381–394.

Rodell, Matthew, Velicogna, I., & Famiglietti, J. S. (2009). Satellite-based estimates of groundwater depletion in India. *Nature*, *460*(7258), 999–1002.

Roxy, M. K., Ritika, K., Terray, P., Murtugudde, R., Ashok, K., & Goswami, B. N. (2015). Drying of Indian subcontinent by rapid Indian Ocean warming and a weakening land-sea thermal gradient. *Nature Communications*, *6:7423*. https://doi.org/10.1038/ncomms8423

Shah, H. L., & Mishra, V. (2016). Hydrologic Changes in Indian Subcontinental River Basins (1901–2012). *Journal of Hydrometeorology*, *17*(10), 2667–2687.

Shah, T. (2009). Climate change and groundwater: India's opportunities for mitigation and adaptation. *Environmental Research Letters*, *4*(3), 035005. https://doi.org/10.1088/1748-9326/4/3/035005

Siebert, S., Henrich, V., Frenken, K., & Burke, J. (2013). Update of the digital global map of irrigation areas to version 5. *Rheinische Friedrich-Wilhelms-Universität, Bonn, Germany and Food and Agriculture Organization of the United Nations, Rome, Italy*.

Silber, J. H., Rosenbaum, P. R., & Ross, R. N. (1995). Comparing the contributions of groups of predictors: which outcomes vary with hospital rather than patient characteristics? *Journal of the American Statistical Association*, *90*(429), 7–18.

Taylor, R. G., Scanlon, B., Döll, P., Rodell, M., Van Beek, R., Wada, Y., et al. (2013). Ground water and climate change. *Nature Climate Change*, *3*(4), 322–329.

Thomas, A. C., Reager, J. T., Famiglietti, J. S., & Rodell, M. (2014). A GRACE-based water storage deficit approach for hydrological drought characterization. *Geophysical Research Letters*, *41*(5), 2014GL059323. https://doi.org/10.1002/2014GL059323

Tiwari, V. M., Wahr, J., Swenson, S. C., Rao, A. D., & Singh, B. (2009). Land water storage variation over Southern India from space gravimetry. In *AGU Fall Meeting Abstracts* (Vol. 1, p. 06). Retrieved from http://adsabs.harvard.edu/abs/2009AGUFM.G33E..06T

Wada, Y., van Beek, L. P. H., van Kempen, C. M., Reckman, J. W. T. M., Vasak, S., & Bierkens, M. F. P. (2010). Global depletion of groundwater resources: GLOBAL GROUNDWATER DEPLETION. *Geophysical Research Letters*, *37*(20), n/a-n/a. https://doi.org/10.1029/2010GL044571

Wada, Y., Beek, Lph., & Bierkens, M. F. (2012). Nonsustainable groundwater sustaining irrigation: A global assessment. *Water Resources Research*, *48*(6), W00L06.

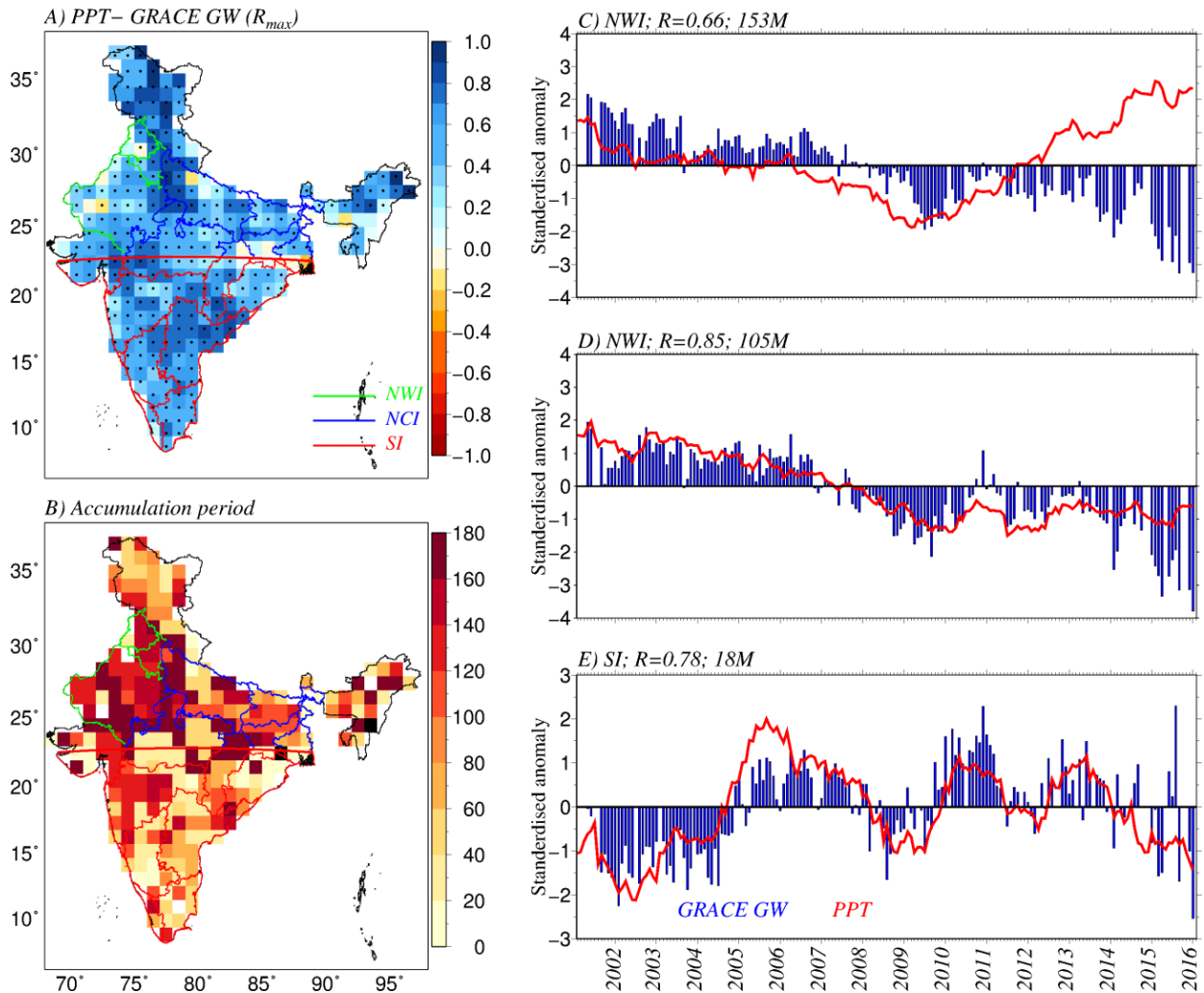

**Figure 1. The linkage between long-term changes in precipitation and groundwater storage anomaly (GWSA) from the GRACE satellites in India** (A) Correlation between GWSA estimated from the GRACE satellites and accumulated precipitation for the selected accumulation period, (B) Optimal period (time-period (1-180 months)) for which correlation between GWSA and accumulated precipitation is the highest, (C-E) Standardized GWSA and precipitation anomalies for the accumulation period for northwest (NWI), north-central (NCI), and south India (SI).

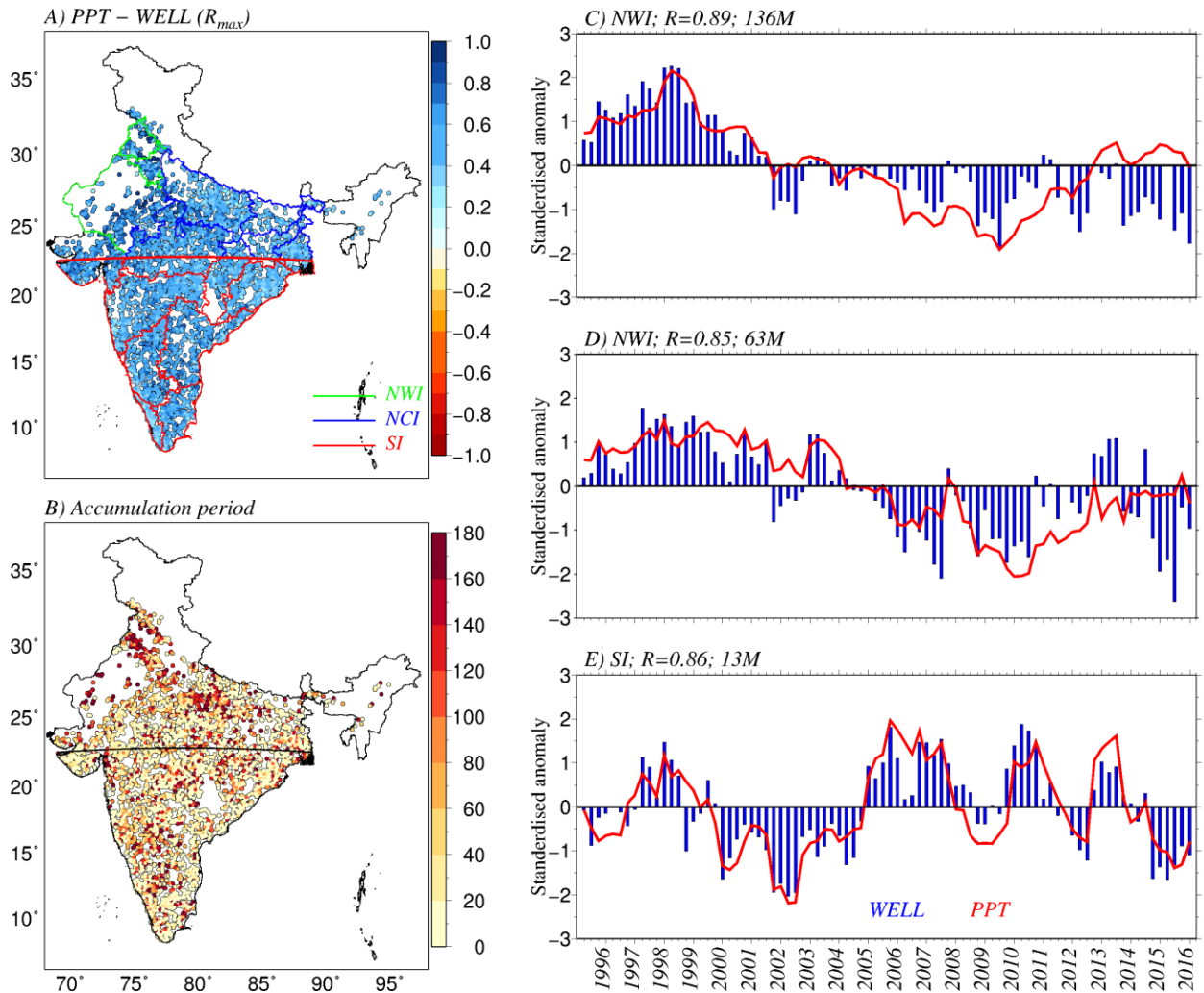

Figure 2. **The linkage between long-term changes in precipitation and groundwater storage anomaly (GWSA) derived from the well level observation in India** (A) Correlation between GWSA estimated from the well level observations and accumulated precipitation for the selected accumulation period, (B) Optimal period (time-period (1-180 months)) for which correlation between GWSA and accumulated precipitation is the highest, (C-E) Standardized GWSA (from well level observations of January, May, September and November) and precipitation anomalies for the accumulation period for northwest (NWI), north-central (NCI), and south India (SI).

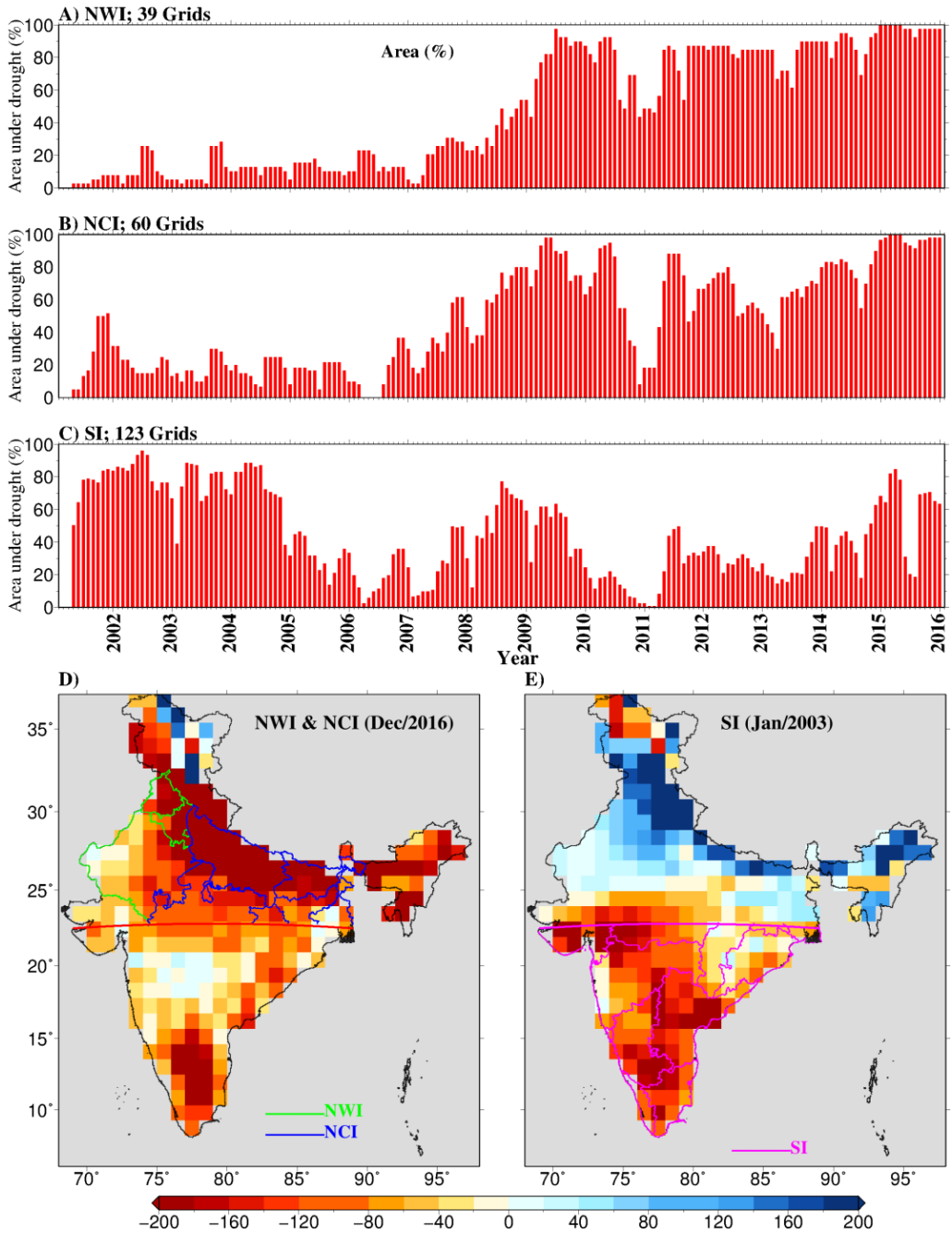

Figure 3. **Persistence of groundwater drought in north India.** (A-C) Area (%) under groundwater drought (red bars) estimated using the GRACE data for each month during 2002-2016. (D) Groundwater storage anomaly (mm) for the highest negative anomaly in region average time series (December, 2016) of groundwater drought in NWI and NCI, and (E) same as (D) but for the SI (January 2003).

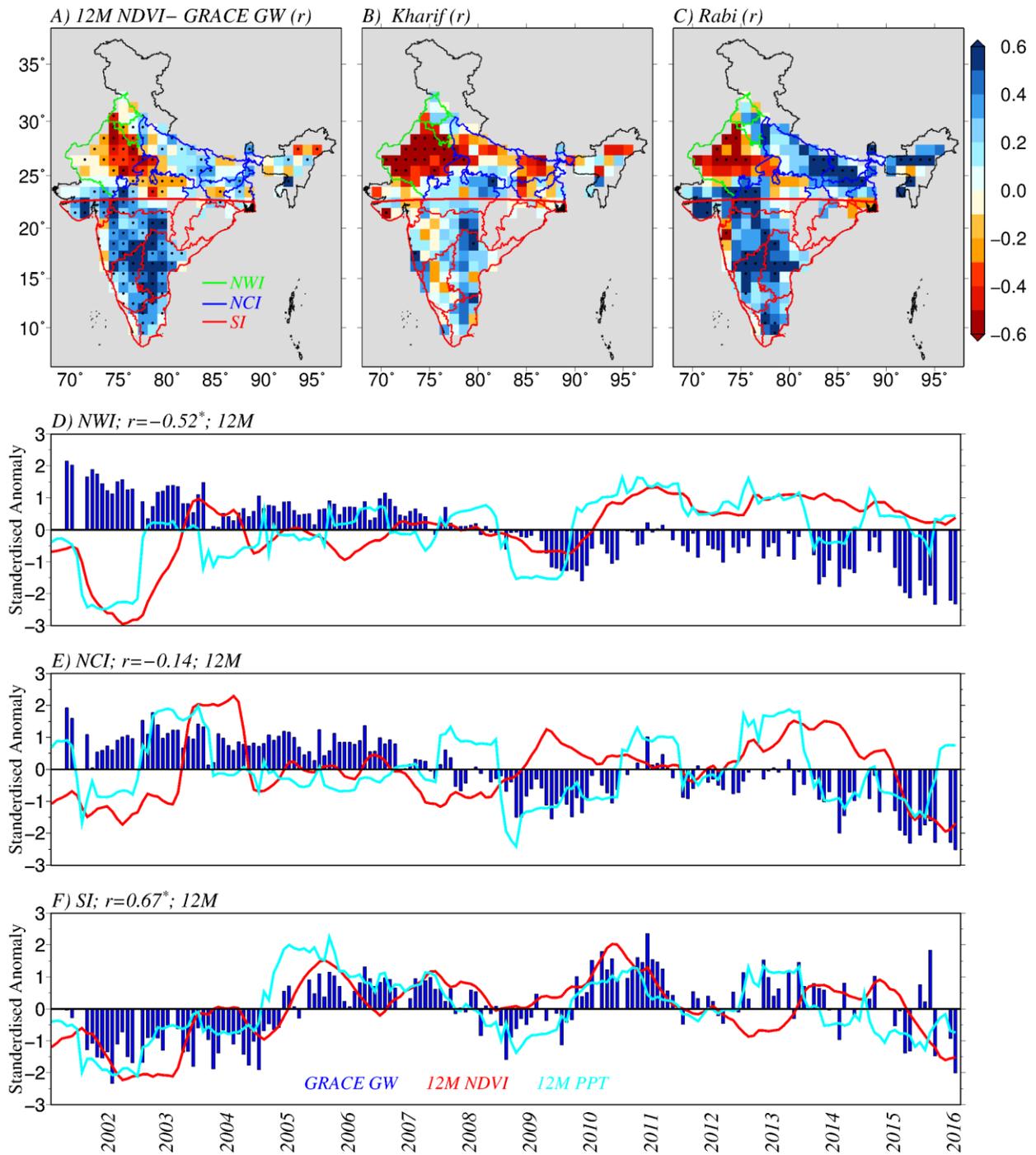

Figure 4. **The role of vegetation on groundwater storage variability in India**. A) correlation between anomalies 12 months accumulated NDVI and GRACE groundwater storage anomaly. B) Correlation between Kharif season (JJAS) mean NDVI and GWSA C) same as (B) but for Rabi season (ONDJFM). (D-F) The standardised anomalies of GWSA and 12 months accumulated NDVI, precipitation for the NWI, NCI, and SI. Correlation (r) values between standardised anomalies of GWSA and 12 months accumulated NDVI are shown. The '*' represents the correlation values are significant at 95 % confidence level.

**Supporting Information**

**Groundwater pumping to increase food production causes persistent groundwater drought in India**

Akarsh Asoka[1] and Vimal Mishra[1]
1. Civil Engineering and Earth Sciences, Indian Institute of Technology (IIT) Gandhinagar, Gujarat, India

Corresponding author: vmishra@iitgn.ac.in

Figures: S1-S6

Tables: S1-S4

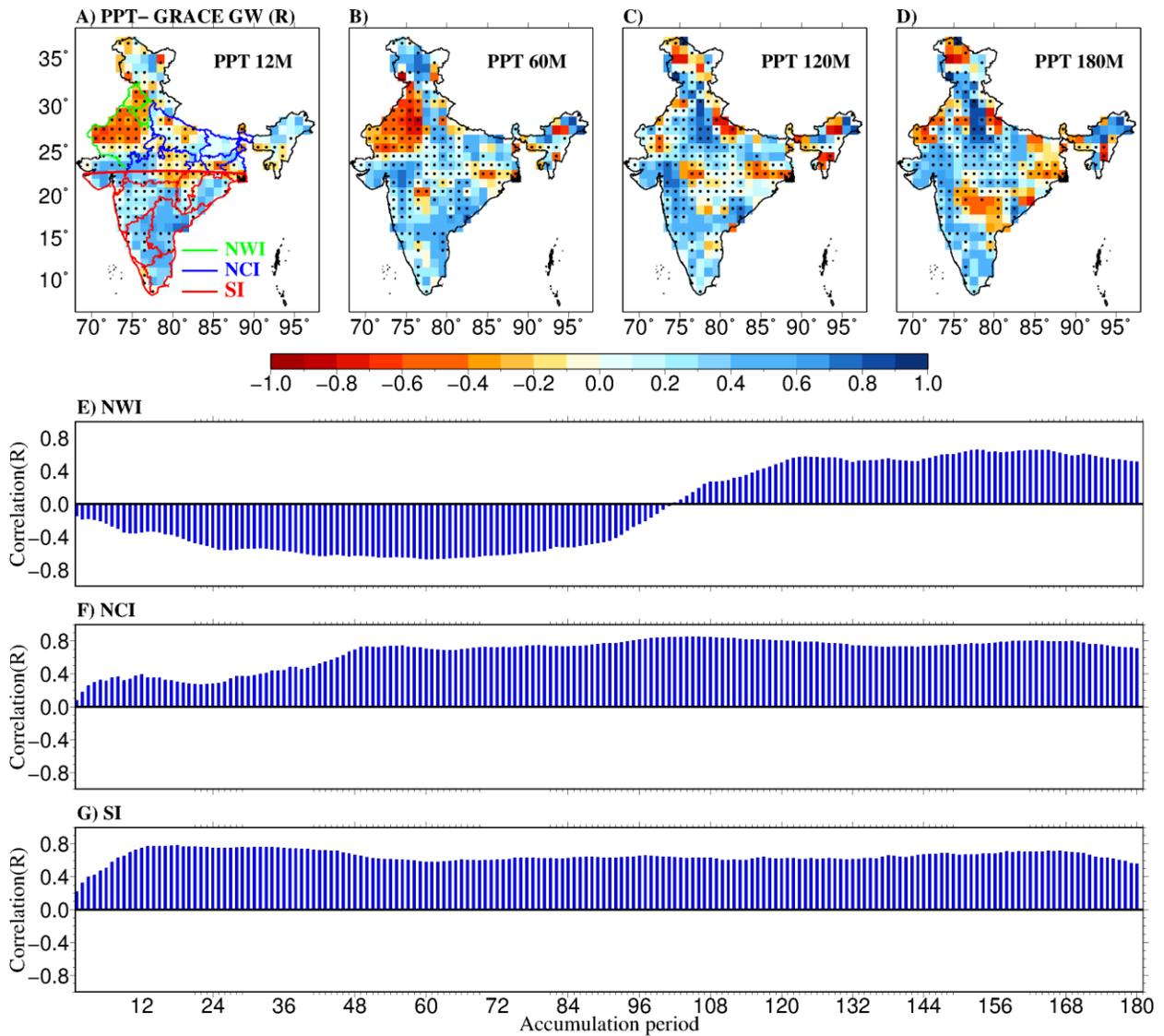

Figure S1. (A-D) The correlation between 12, 60, 120 and 180 month accumulated precipitation anomaly and GWSA. (E-G) The correlation between 1-180 month accumulated precipitation anomaly and GWSA in the NWI, NCI and SI.

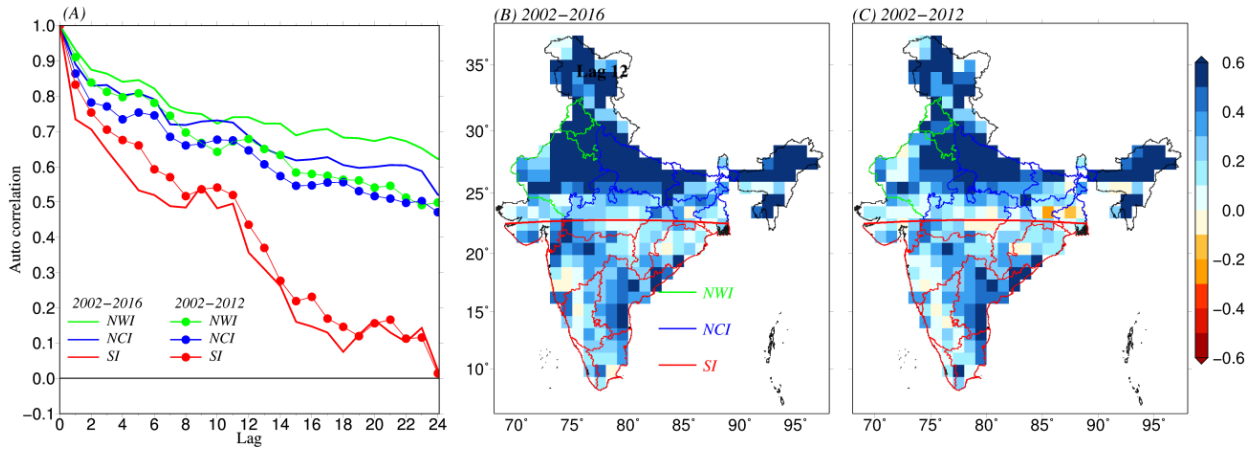

Figure S2. (A) The autocorrelation of standardized GWSA for NWI, NCI and SI during 2002-2016 and 2002-2012 for 0 to 24-month lag. The autocorrelation of standardized GWSA for India at 12-month lag during (B) 2002-2016 and (C) 2002-2012

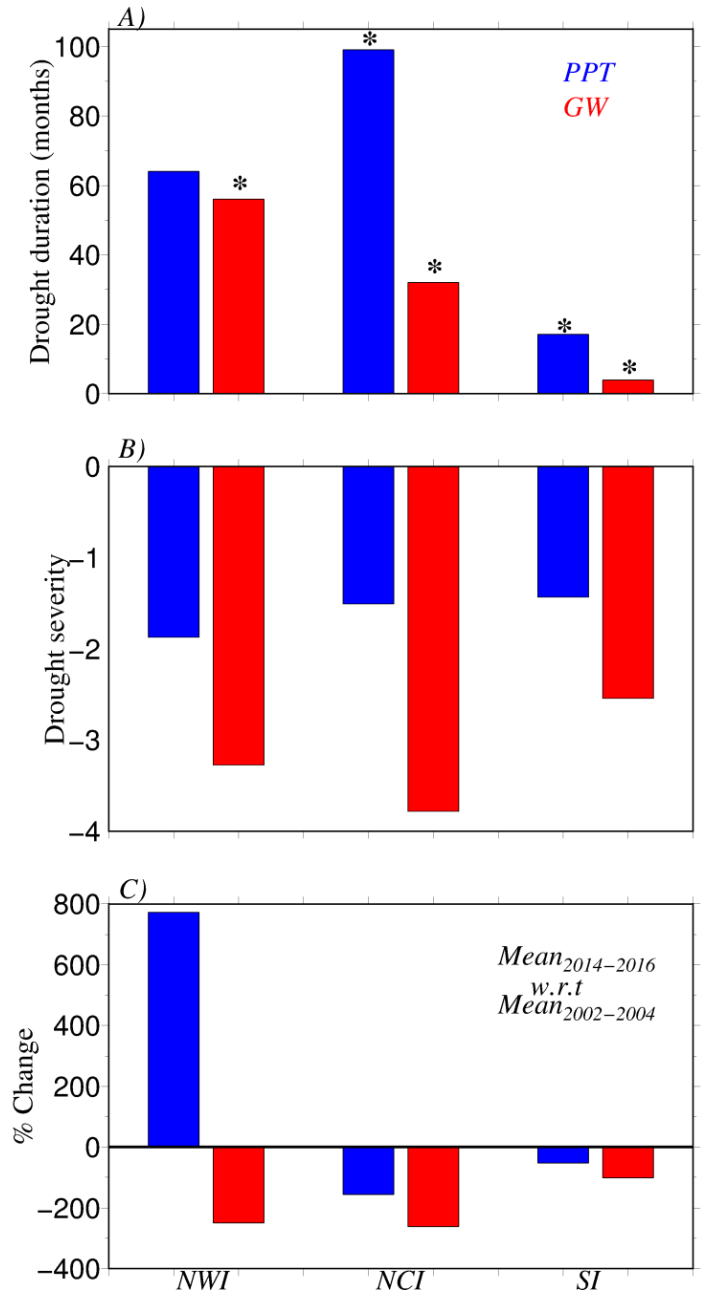

Figure S3. A) The duration of the drought (months) for the latest drought estimated from GRACE groundwater storage anomaly (red bar) and optimal accumulated precipitation (blue bar) for NWI, NCI and SI. The '*' shows the drought is persisting at the end of 2016. B) Same as A) but the drought severity (maxumum negative anomaly) of the latest drought. C) The percentage change in groundwater storage anomaly and optimal accumulated precipitation anomaly in 2014- 2016 (last 3 year mean) with respect to 2002-2004 mean (initial 3 yar mean).

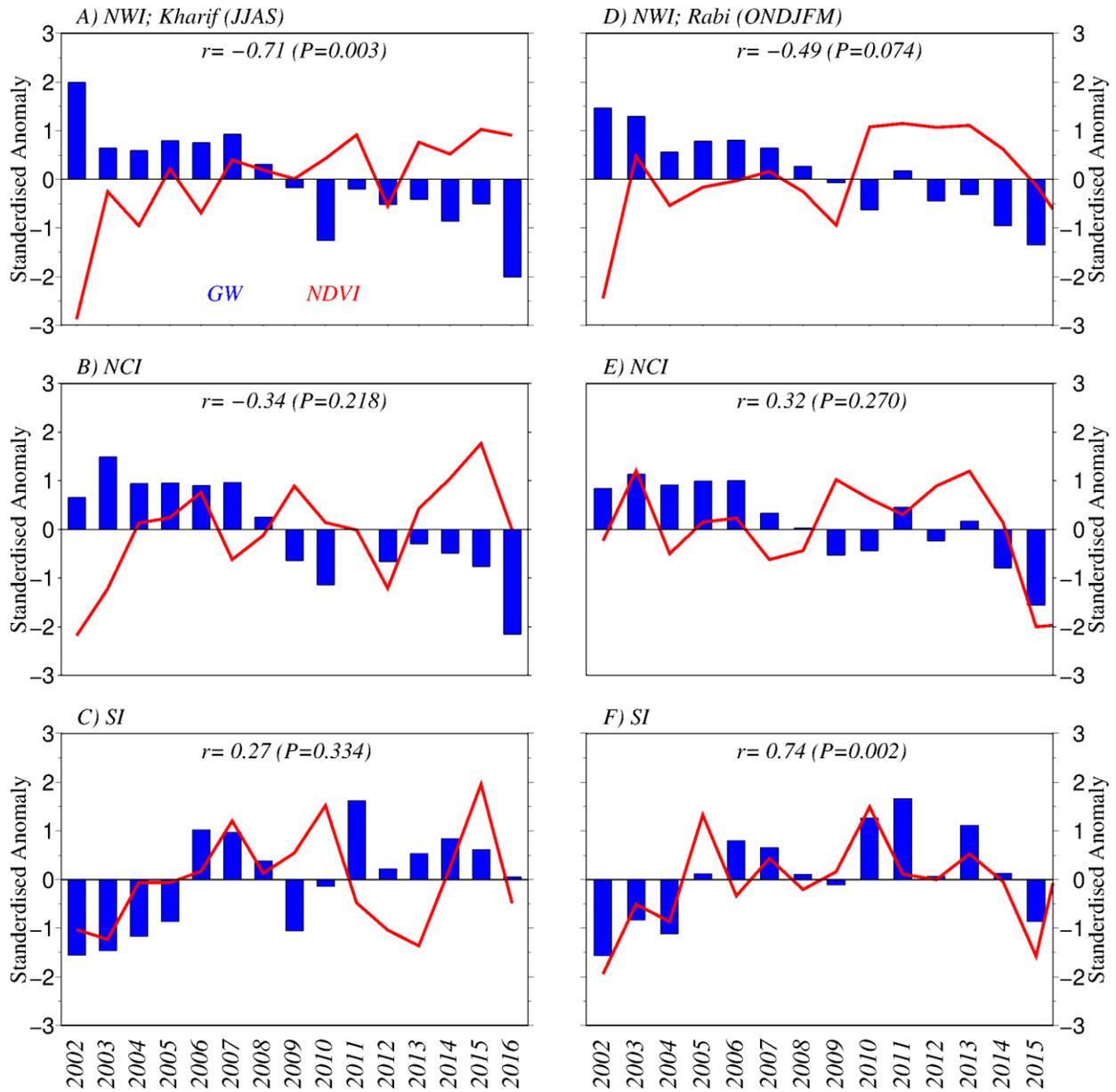

Figure S4. (A-C) The standardized anomalies of GWSA, NDVI and LST in the groundwater irrigated area during kharif season (JJAS) for NWI, NCI and SI from 2002-2016. (D-F) Same as (A-C) but for the rabi season (ONDJFM) during 2002-2015. Considered only the NDVI pixels with more than 60% area is irrigated with groundwater (based on FAO percentage area irrigated with groundwater and such pixels are more than 50 % of the 1º x 1º GRACE grid)

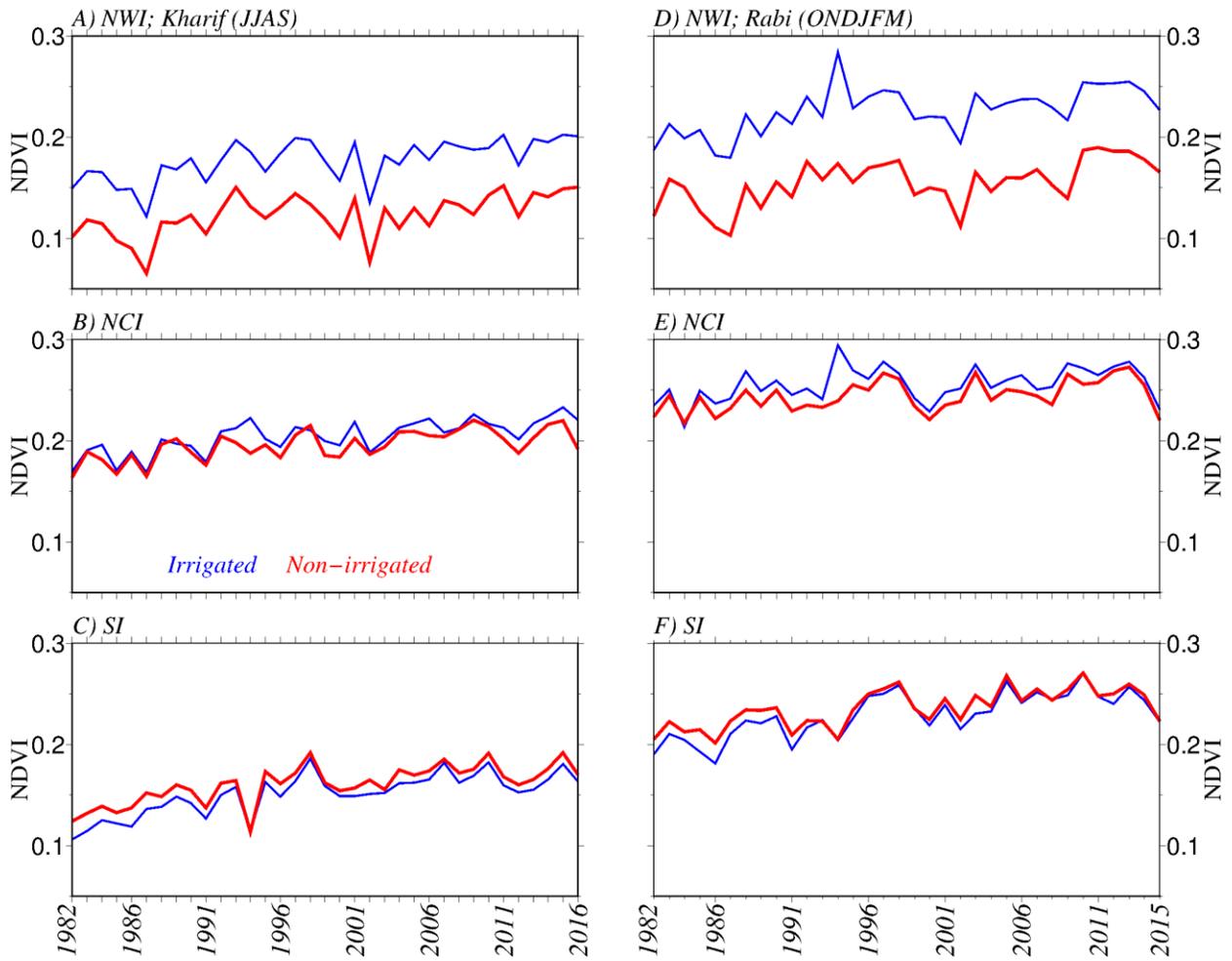

Figure S5. (A-C) The irrigated and non-irrigated area time series of NDVI in the NWI, NCI and SI during kharif season (JJAS) from 1982-2016. (D-F) same as (A-C) but for the rabi season (ONDJFM) from 1982-2015.

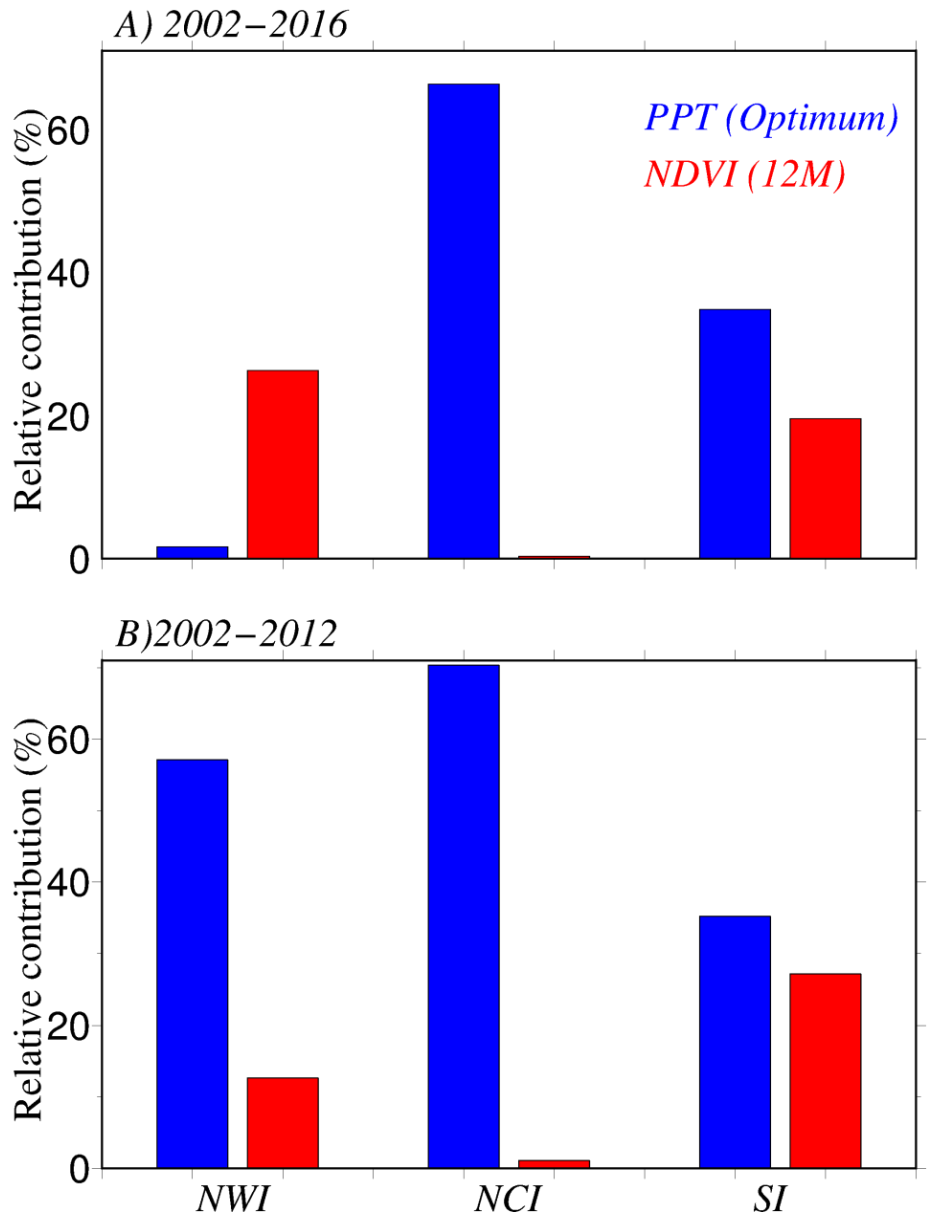

Figure S6. Relative importance ($R^2$ in %) of 12-month NDVI and optimum precipitation [153 (NWI), 105 (NCI) and 18 (SI)] on GWSA for 2002-2016 and 2002-2012 at 95% confidence level

**Table S1. The optimal accumulation period of precipitation for NWI, NCI and SI based on median correlation analysis and single time analysis.**

|  |  | Median |  |  |  | Full time series |  |  |
|---|---|---|---|---|---|---|---|---|
| Data | Region | R | P-value | Optimum Accumulation period | Standard Deviation of R | R | P-value | Optimum Accumulation period |
| GRACE-PPT | NWI | 0.67 | 0.00 | 153 | 0.26 | 0.09 | 0.25 | 177 |
|  | NCI | 0.84 | 0.00 | 105 | 0.18 | 0.84 | 0.00 | 170 |
|  | SI | 0.78 | 0.00 | 18 | 0.04 | 0.73 | 0.00 | 18 |
|  |  |  |  |  |  |  |  |  |
| WELL-PPT | NWI | 0.89 | 0.00 | 136 | 0.04 | 0.83 | 0.00 | 167 |
|  | NCI | 0.85 | 0.00 | 63 | 0.05 | 0.78 | 0.00 | 26 |
|  | SI | 0.86 | 0.00 | 13 | 0.01 | 0.86 | 0.00 | 13 |

**Table S2. The summary of groundwater drought in NWI, NCI and SI.**

| Location | Duration of latest drought event (months) | | Duration of longest drought event based on GWSA (months) | Maximum departure in GWSA (mm) | |
|---|---|---|---|---|---|
|  | Optimal accumulated precipitation | GWSA |  | Positive/ wettest | Negative/ driest |
| NWI | 64 (04/2007 to 08/2012) | 56 (04/2012 to 12/2016) | 56 (04/2012 to 12/2016) | 76.76 (09/2002) | -103.70 (12/2016) |
| NCI | 99 (09/2008 to 12/2016) | 32 (04/2014 to 12/2016) | 32 (04/2014 to 12/2016) | 112.31 (05/2002) | -205.55 (12/2016) |
| SI | 17 (07/2015 to 12/2016) | 4 (08/2016 to 12/2016 | 21 (04/2002 to 12/2003) & 02/2004 to 10/2005) | 90.56 (11/2011) | -108.50 (01/2003) |

**Table S3. Correlation between 12-month NDVI and GWSA in NWI, NCI and SI**

| | Median | | Full time series | | |
|---|---|---|---|---|---|
| Region | r-value | p-value | r-value | p-value | Standard deviation in r |
| NWI | -0.52 | 0.00 | -0.52 | 0.00 | 0.07 |
| NCI | -0.14 | 0.18 | -0.01 | 0.94 | 0.11 |
| SI | 0.67 | 0.00 | 0.61 | 0.00 | 0.03 |

Table S4. Relative importance of 4,12 and 24 Month NDVI and optimum precipitation [153 (NWI), 105 (NCI) and 18 (SI)] on GWSA for 2002-2016 and 2002-2012 at 95% confidence level

| Time period | Region | Relative Importance ($R^2$) | | Lower bound ($R^2$) | | Upper bound ($R^2$) | | Model $R^2$ |
|---|---|---|---|---|---|---|---|---|
| | | PPT | NDVI | PPT | NDVI | PPT | NDVI | |
| 2002-2016 (04M NDVI) | NWI | 0.02 | 0.13 | 0.00 | 0.03 | 0.11 | 0.26 | 0.15 |
| | NCI | 0.66 | 0.01 | 0.59 | 0.00 | 0.73 | 0.05 | 0.68 |
| | SI | 0.42 | 0.12 | 0.34 | 0.06 | 0.50 | 0.19 | 0.53 |
| 2002-2012 (04M NDVI) | NWI | 0.53 | 0.07 | 0.38 | 0.01 | 0.67 | 0.18 | 0.61 |
| | NCI | 0.71 | 0.00 | 0.63 | 0.00 | 0.78 | 0.04 | 0.72 |
| | SI | 0.45 | 0.15 | 0.38 | 0.08 | 0.53 | 0.23 | 0.60 |
| 2002-2016 (12M NDVI) | NWI | 0.02 | 0.26 | 0.00 | 0.16 | 0.10 | 0.38 | 0.28 |
| | NCI | 0.66 | 0.00 | 0.58 | 0.00 | 0.73 | 0.03 | 0.67 |
| | SI | 0.35 | 0.20 | 0.27 | 0.13 | 0.43 | 0.27 | 0.54 |
| 2002-2012 (12M NDVI) | NWI | 0.57 | 0.13 | 0.44 | 0.06 | 0.68 | 0.22 | 0.70 |
| | NCI | 0.70 | 0.01 | 0.61 | 0.00 | 0.77 | 0.05 | 0.71 |
| | SI | 0.35 | 0.27 | 0.29 | 0.21 | 0.42 | 0.34 | 0.62 |
| 2002-2016 (24M NDVI) | NWI | 0.01 | 0.42 | 0.00 | 0.32 | 0.08 | 0.52 | 0.43 |
| | NCI | 0.65 | 0.02 | 0.57 | 0.01 | 0.71 | 0.06 | 0.67 |
| | SI | 0.35 | 0.24 | 0.28 | 0.17 | 0.43 | 0.31 | 0.58 |
| 2002-2012 (24M NDVI) | NWI | 0.53 | 0.18 | 0.42 | 0.10 | 0.63 | 0.28 | 0.71 |
| | NCI | 0.70 | 0.02 | 0.62 | 0.00 | 0.78 | 0.05 | 0.72 |
| | SI | 0.35 | 0.32 | 0.29 | 0.25 | 0.41 | 0.39 | 0.67 |